\documentclass[12pt]{article}
\usepackage{latexsym}
\usepackage{amsmath}
\usepackage{amsfonts}
\usepackage{amssymb}
\usepackage{color}

\newcommand{\Be}{\begin{equation}}
\newcommand{\Ee}{\end{equation}}

\title{Free Particle Wave Function and Niederer's Transformation}
\author{K. Andrzejewski,
J. Gonera\thanks{e-mail:jgonera@uni.lodz.pl}, P. Kosi\'nski\\\\
\small Department of Theoretical Physics and Computer Science, \\
\small University of \L\'od\'z,\\
\small Pomorska 149/153, 90-236 {\L}\'od\'z, Poland
}
\date{}
\begin{document}

\maketitle 
\begin{abstract}
The solutions to the free Schroedinger equation discussed by P. Strange (arXiv: 1309.6753) and A. Aiello (arXiv: 1309.7899) are analyzed. It is shown that their properties can be explained with the help of Niederer's transformation.
\end{abstract}
\section{Introduction}
\par
Recently Strange \cite{b1} considered a specific solution to one-dimensional free particle Schroedinger equation in which the space and 
time dependence are not separable. The solution exhibits few peaks in probability density which accelerate with time. It has been 
analyzed in \cite{b1} quantum and semiclassical regimes along the same lines as the Airy accelerating solution constructed by Berry and Balazs \cite{b2}. Strange's solution was further generalized to two dimensions by Aiello \cite{b3}.
\par
In the present short note we show that the form and properties of such solution are direct consequences of 
Niederer's transformation \cite{b4} relating, both on classical and quantum levels, the harmonic oscillator and the free particle of the 
same mass. We generalize the results presented in \cite{b1} concerning the behavior of probability density and its relation to the properties of family of classical trajectories.
\par
It is also shown that the solutions under consideration can be easily generalized to any dimension.

\section{Niederer's transformation}
\par
Niederer \cite{b4} (see also \cite{b5}) constructed the mapping which transforms the harmonic oscillator motion into the free one. \\
Given a $d$-dimensional oscillator described by the Lagrangian

\Be
\label{f1}
L = \frac{m}{2} \dot{\vec{x} }^2 - \frac{m\omega ^2}{2}\vec{x} ^2
\Ee
we consider the following point transformation 

\begin{eqnarray}
\label{f2}
\tau& = \frac{1}{\omega } \tan{(\omega t)},\qquad t=\frac{1}{\omega } \arctan{(\omega \tau )}\\
\vec{y}&= \frac{\vec{x}}{\cos{(\omega t)}}, \qquad \qquad \vec{x} = \vec{y} (1+ \omega ^2\tau ^2)^{-\frac{1}{2}}\nonumber 
\end{eqnarray}
One easily checks the following identity

\Be
\label{f3}
Ldt = \frac{m}{2} \big( \frac{d\vec{y} }{d\tau }\big) ^2 d\tau  - d\big(\frac{m\omega}{4} \sin {(2\omega t)}\vec{y} ^2 \big)
\Ee
which tells us that $\vec{y} (\tau )$ describes free motion provided $\vec{x}(t)$ obeys harmonic oscillator equation of motion. In 
particular, the mapping (\ref{f2}) transforms the harmonic motion in the interval $ (-\frac{\pi }{2\omega },\frac{\pi }{2\omega })$
(half of the period) into free dynamics for $-\infty < t< \infty $.
\par
Transformation described by eqs. (\ref{f2}) has its quantum counterpart. Namely, if $\psi (\vec{x},t)$ obeys the Schroedinger equation for harmonic oscillator then 
\Be
\label{f4}
\chi (\vec{y},\tau ) = (1 + \omega ^2\tau ^2)^{-\frac{1}{4}}e^{\frac{im\omega ^2\tau }{2(1+\omega ^2\tau )}\vec{y} ^2}\psi (\vec{y}(1+\omega ^2\tau ^2)^{-\frac{1}{2}},\frac{1}{\omega }\arctan{(\omega \tau )} )
\Ee
is a solution to the free Schroedinger equation. The structure of the relation given by eq. (\ref{f4}) is transparent. First, the 
arguments of the wave function are replaced by the appropriate functions of new ones according to the classical formulae (\ref{f2}); 
then two factors are added: the first one accounts for proper normalization while the second is related to the fact that under the transformation (\ref{f2}) the Lagrangian transforms by a total derivative (cf. eq. (\ref{f3})).
\par
The inverse transformation reads:

\Be
\label{f5}
\psi (\vec{x},t) = (\cos{(\omega t)})^{-\frac{1}{2}} e ^{-\frac{im\omega }{2} \tan{(\omega t)}\vec{x}^2}\chi (\frac{\vec{x}}{\cos{(\omega t)}},\frac{1}{\omega }\tan{\omega t})        
\Ee
Taking $d=1$ and $\psi (x,t)$ as the eigenfunction of harmonic oscillator Hamiltonian

\Be
\label{f6}
\psi (x,t) = \frac{1}{\sqrt{2^n n!}}\big( \frac{m\omega }{\pi}\big) ^{\frac{1}{4}} e^{-i\omega (n+ \frac{1}{2})t} e ^{-\frac{m\omega x^2}{2}} H_n \big( (m\omega )^{\frac{1}{2}} x\big)
\Ee
one arrives at the formula (\ref{f2}) from the Strange paper \cite{b1} provided the identification $\omega = \frac{1}{t_c}$ has been made.
\par
One can also easily  construct the multidimensional solutions. For example, let us take $d=2$ and let $\psi (\vec{x},t)$ be the common 
eigenfunction of energy and angular momentum; up to a normalization factor 

\Be
\label{f7}
\psi _{n,l}(\vec{x},t) = e^{-i\omega (n+\frac{|l|}{2} + \frac{1}{2})t}e^{-\frac{m\omega r^2}{2}}r^{|l|} F(-n,|l| + 1, m\omega r^2)
\Ee
Applying eq. (\ref{f4}) to $\psi_{0,l}$ one arrives at the wave function considered by Aiello \cite{b3}. In a similar way one can construct
the three  and higher dimensional examples.
\par
Let us now come back to the one-dimensional case. We would like to analyze the behavior of probability density for the solutions under 
consideration. Let $\psi _n(x,t)$ be $n$-th stationary solution. Then eq.(\ref{f4}) implies

\Be
\label{f8}
|\chi (y,\tau )|^2 = (1 + \omega ^2 \tau ^2)^{-\frac{1}{2}}\rho _n(y(1+\omega ^2\tau ^2)^{-\frac{1}{2}}),  
\Ee
$\rho _n(x)$ being time-dependent probability density for stationary state of harmonic oscillator. Now, $\rho _n(x)$ has $n + 1$ maxima. 
Therefore, the probability density $|\chi (y, \tau )|^2$ has $n + 1$ peaks $y_n(\tau )$ which travel with time along hyperbolic  
trajectories

\Be
\label{f9}
y_n(\tau ) = y_n(0) (1+\omega ^2\tau ^2)^{\frac{1}{2}}
\Ee
and broaden according to $(1 + \omega ^2 \tau ^2)^{\frac{1}{2}}$ law. This generalizes to any $n$ the findings of Ref. \cite{b1} for $n=2$.
\par
Berry and Balazs \cite{b2}  found the solution to the free Schroedinger equation in form of wave packet propagating in space without
distortion and with constant acceleration. Although being not square integrable it does not provide the counterexample to Ehrenfest theorem the 
appearance of accelerated motion for free particle is slightly disturbing. Therefore, Berry and Balazs proposed an explanation of this fact
based on considering the family of classical trajectories related, via semiclassical  analysis, to the wave function under 
consideration; the acceleration of the packet is ascribed to the curvature of envelope of this family.
\par
In the case of the solution considered by Strange, which is normalizable, the Ehrenfest theorem cannot be broken. However, the 
probability density posses several peaks which travel with acceleration. As shown by Strange in $n=2$ case their acceleration can be 
also related to the  behavior of some family of classical trajectories. The pictures based on Niederer's transformation allows us to 
give simple interpretation for any $n$. The $n$-th eigenstate of the harmonic oscillator Hamiltonian corresponds to the family of 
classical trajectories parametrized by the angle $\alpha $

\Be
\label{f10}
x(t,\alpha ) = \sqrt{\frac{2E_n}{m\omega ^2}} \cos{(\omega t + \alpha )}
\Ee
which, in view of eq. (\ref{f2}) yield the family of free trajectories

\Be
\label{f11}
y( \tau ,\alpha ) = \sqrt{\frac{2E_n}{m\omega ^2}} (\cos \alpha - \omega \tau \sin{\alpha }  )
\Ee
The envelope of this family is obtained by solving

\Be
\label{f12}
\frac{\partial y(\tau ,\alpha )}{\partial \alpha } = 0
\Ee
with respect to $\alpha $ and reinserting into eq. (\ref{f11}). This yields

\Be
\label{f13}
y(\tau ) = \pm \sqrt{\frac{2E_n}{m\omega ^2}} (1 + \omega ^2\tau ^2)^{\frac{1}{2}}
\Ee
Now, $ x_{\pm}  = \pm \sqrt{\frac{2E_n}{m\omega ^2}} $ are the classical turning points restricting the classically allowed region for 
the harmonic oscillator motion with the energy $E_n$. In the semiclassical regime, $n  \gg 1$, $n\hbar$ fixed, the extreme maxima of 
probability density are placed at points differing by $O(\frac{1}{n})$ distance from turning points. Therefore, eq. (\ref{f4}) tells us 
that eq.(\ref{f13}) describes in the semiclassical limit the motion of extreme maxima of probability density for the corresponding free 
Schroedinger equation.

\section{Conclusion}
\par
We have shown that the "exotic" solutions to the free Schroedinger equation considered recently \cite{b1}, \cite{b3} can be easily generated
by using Niederer's transformation \cite{b4}, \cite{b5}. Their slightly peculiar properties are naturally explained in terms of this transformation.
\par
We considered here only the solutions to the free Schroedinger equation, which are the images, under the Niderer transformation, of 
stationary states of harmonic oscillators. However, there are also other interesting states of harmonic oscillator which could generate 
some nice  free wave functions. For example, coherent states yield the Gaussian packet of constant width oscillating according to the
classical equations of motion. This and other similar cases will be studied elsewhere.

\par
{\bf Acknowledgements}
The authors are grateful to Professors Pawe\l \, Ma\'slanka and Cezary Gonera
for useful remarks and discussion. The first author is supported in part from the earmarked
subsidies MNiSzW for Young Scientists, as well  as laureate of the  Uiversity of Lodz Foundation prize.


\begin{thebibliography}{99}
\bibitem{b1}
P. Strange, {\it Semiclassical and Quantum Analysis of a Free Particle Herimite Wavefunction}, arXiv: 1309.6753
\bibitem{b2}
M. V. Berry, N. L. Balazs, Am. J. Phys {\bf 47} (1979), 264
\bibitem{b3}
A. Aiello, {\it Comment on "Semiclassical and Quantum Analysis of a Focusing Free Particle Herimite Wavefunctions, by PPaul Strange (arXiv: 1309.6753)}", arXiv: 1309.7899
\bibitem{b4}
U. Niederer, Helv. Phys. Acta {\bf 46} (1973), 191
\bibitem{b5}
G. Burdet, C. Duval, M. Perrin, Lett. Math. Phys. {\bf 10} (1985), 255\\
P-M. Zhang, P. A. Horvathy, Phys Lett. {\bf B 702} (2011), 177\\
P. A. Horvathy, J.-C. Yera, Int. Journ. Theor. Phys. 48 (2009), 3139\\
A. V. Galajinsky, Nucl. Phys. {\bf B 832} (2010), 586\\
A. V. Galajinsky, I. Masterov, Nucl. Phys. {\bf B 866} (2013), 212\\
A. V. Galajinsky. I. Masterov, Phys. Lett. {\bf B 702} (2011), 265
\end{thebibliography}
\end{document}